# Two source emission behaviour of alpha fragments of projectile having energy around 1 GeV per nucleon

V. Singh
*Nuclear and Astroparticle Physics Lab., Physics Department, Banaras Hindu University, Varanasi 221 005, INDIA*

M. K. Singh, Ramji Pathak
*Department of Physics, Tilakdhari Postgraduate College, Jaunpur 222 002, INDIA*

The emission of projectile fragments alpha has been studied in $^{84}$Kr interactions with nuclei of the nuclear emulsion detector composition at relativistic energy below 2 GeV per nucleon. The angular distribution of projectile fragments alpha in terms of transverse momentum could not be explained by a straight and clean-cut collision geometry hypothesis of Participant – Spectator (PS) Model. Therefore, it is assumed that projectile fragments alpha were produced from two separate sources that belong to the projectile spectator region differing drastically in their temperatures. It has been clearly observed that the emission of projectile fragments alpha are from two different sources. The contribution of projectile fragments alpha from contact layer or hot source is a few percent of the total emission of projectile fragments alphas. Most of the projectile fragments alphas are emitted from the cold source. It has been noticed that the temperature of hot and cold regions are dependent on the projectile mass number.

## 1. INTRODUCTION

Nuclear fragmentation is an important experimental phenomenon in nucleus – nucleus collisions at relativistic high energy and heavy ion [1]. The PS Model is a simple and basic model for the study of the high energy nucleus – nucleus collisions. The emission of projectile fragments alpha has been studied in lots of high energy heavy ion experiments having energy more than a GeV per nucleon and the general consensus was that they are coming out from the projectile spectator part in the context of a PS Model [2]. According to the PS Model, the interacting system in high energy nucleus – nucleus collisions can be clearly divided into three regions such as target and projectile spectators and a participant region. The overlapping region of the two colliding nuclei is called the participant region and the other regions are called the target and the projectile spectators, respectively.

Data from lighter projectiles has been successfully explained with the help of PS Model but data from heavy ion interactions pointed finger towards the PS Model because experiments observed many projectile fragments alpha having larger emission angle that could not be explained by the straight and clean-cut collision geometry picture of the PS Model. Therefore, this large angle scattering indicated the failure of the PS Model and forced us to rethink and modify this model according to the physics requirements and the idea of multiple sources i.e. emission of projectile fragments alpha from different sources having different temperature has been introduced [3]. In this paper, we analyzed the transverse momentum distributions of projectile fragments alpha produced in $^{84}$Kr interactions with the nuclei of the Nuclear Emulsion Detector's (NED) target at around 1 GeV per nucleon kinetic energy [4]. Experimental techniques are briefly described in section 3 while required modification in PS Model is mentioned in section 2. Observed results, which explain data of projectile fragmentation using a two source PS Model, are discussed in section 4.

## 2. THE MODEL

The model used in this paper is based on the two – source emission of projectile fragments assumption. According to this model, the lighter nuclear fragments produced in a projectile spectator have a two emission sources: a hot and a cold source. The hot source is the contact layer of the spectator very close to the participant region and having high excitation degree. The cold source is the other part of the spectator and having low excitation degree. This model has been explained in Ref. [5]. During the collisions, due to the existence of the relative motion between the participant and the spectator regions, the friction is assumed to be caused on the contact layer. In this situation, both the participant and the spectator get the heat due to friction. It takes some time when the contact layer transmits the heat of friction to the rest part of the spectator and therefore, we believe, that this may be the cause of temperature gradient in the spectator region of projectile. The contact layer and the rest part are separated from each other because of the heat of friction. Therefore, the contact layer and the rest part of the spectator are considered as two sources to emit nuclear fragments with two different temperatures. It could be possible that during the collision contact layer portion have highest temperature after participant region. The fall in temperature is rapid towards the farther side of projectile spectator region. The change in temperature follows exponential decay nature and it could be explained from the charge spectrum of the projectile fragments. Considering clear-cut collision geometry of PS Model, it can also be possible that the temperature is almost constant in a layer and the thickness of layers increases with distance from the contact layer as shown in Figure 1 [6].

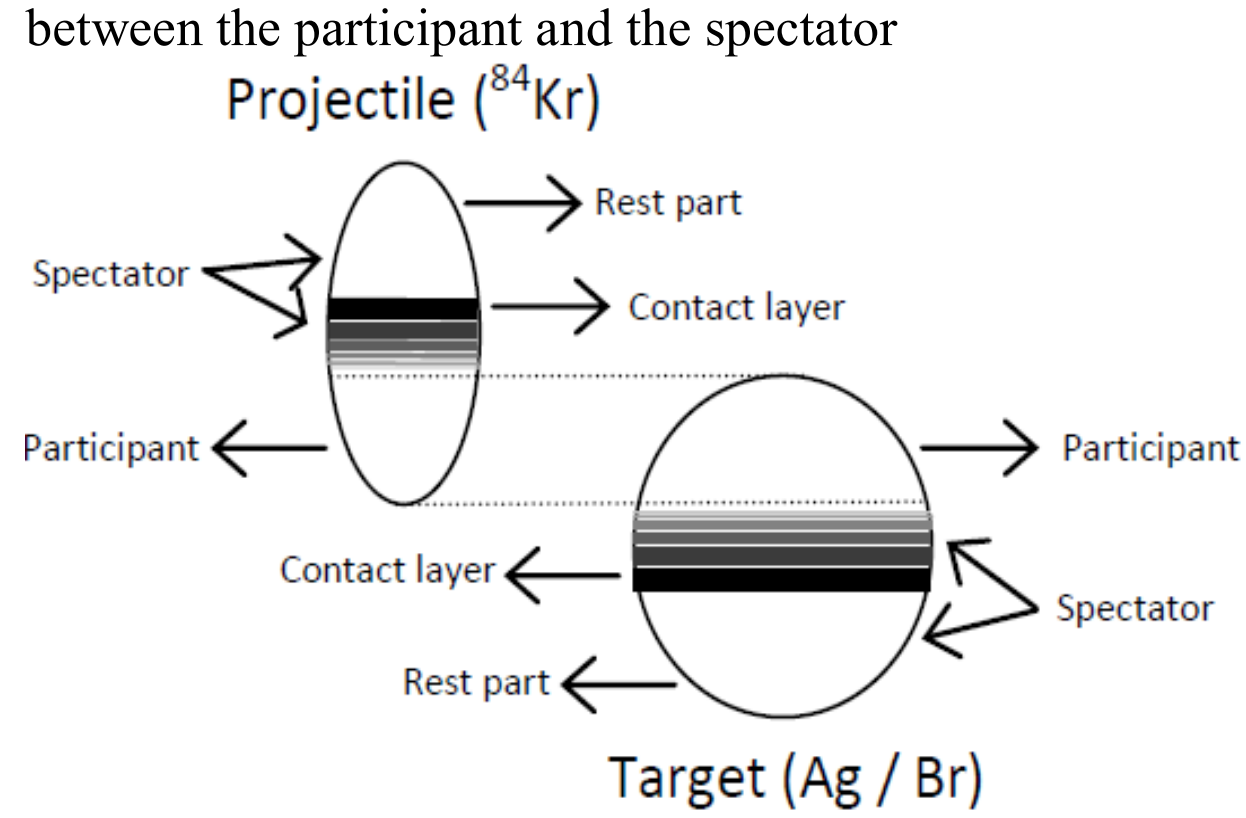

Figure 1: Schematic overview of the contact layer including fine layers of temperatures and the rest part of the projectile and target spectators across the incoming direction of the projectile. Darker lines representing lower temperature regions. The projectile is approaching toward the target and target is at rest.

## 3. EXPERIMENTAL DETAIL

This experiment has been performed in a stack of NIKFI BR-2 Nuclear Emulsion Detectors (NED) having 600 µm thickness exposed at GSI, Darmstadt in Germany to a beam of $^{84}$Kr nuclei at around 1 A GeV [7]. Complete measurements have been done at Banaras Hindu University by the scanning of NED volume using both standard methods [8], having dimension 9.8 cm × 9.8 cm × 0.06 cm, with the help of Olympus BH-2 transmitted light-binocular microscope under 100X oil emersion objective and 15X eyepieces. A total of 600 inelastic interactions of $^{84}$Kr nuclei have been picked up and the charge and angle measurements of projectile fragments of these events have been done. Grain, blob and hole density, and delta ray counting measurements have been done for the estimation of the charge of the light projectile fragments with the accuracy of a unit charge and these methods have been explained in detail in Ref. [9]. For angle measurement of the projectile fragment a small but very effective device called "Gonio-meter" has been used having a least count better than a quarter of a degree.

## 4. RESULTS AND DISCUSSION

The transverse momentum distributions of projectile fragments alpha produced in the nucleus – nucleus collisions at relativistic energy have been analyzed by using a two emission source model [5]. In the NED based experiment, it is not possible to make direct momentum measurement of charge particles / projectile fragments. However, the transverse momentum can indirectly be calculated by using the fact that the fragments have nearly the same momentum per nucleon as that of the projectile in case of fixed target experiment, because when a projectile nucleus with relativistic energy collides with a target nucleus the projectile fragments emitted retain more or less the same momentum per nucleon as of the of two source of projectile fragments alpha emission which is the sum of the two Rayleigh scattering distributions.

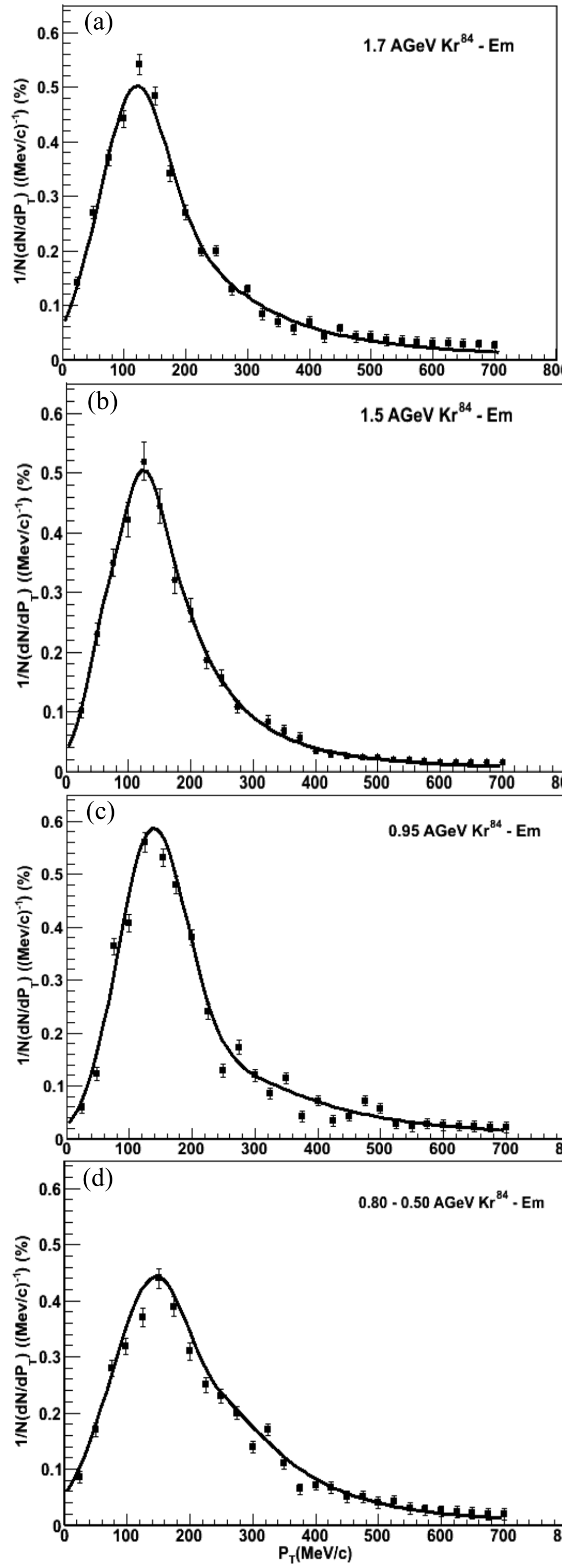

Figure 2: Transverse momentum distribution of projectile fragments alpha emitted in $^{84}$Kr nuclei interactions with emulsion target nuclei at (a) 1.7 A GeV [10], (b) 1.5 A GeV [4], (c) 0.95 A GeV [4] and (d) 0.80-0.50 A GeV [4], respectively. The closed circles are observed values and the solid curve is the calculated values of the assumption

projectile nucleon. So if $p_o$ is the momentum of the incident projectile, the transverse momentum of the fragment of charge Z can be calculated by using the following relation:

$$p_T = A_F\, p_0\, \mathrm{Sin}\theta, \quad (1)$$

where $A_F$ is the mass number of the fragments and θ is the emission angle of the fragments with respect to the projectile direction. Therefore, the pseudo-transverse momentum can be obtained from the measurement of the emission angles.

In order to test the two source fragment emission picture, we compare the sum of two Rayleigh scattering distributions of the transverse momentum with the observed data for similar projectile but with different kinetic energies. The transverse momentum distribution of projectile fragments alpha has been plotted for different kinetic energies and a solid curve is the sum of two Rayleigh scattering distribution function as described in Eq. (12) [5], are superimposed in Figure 2 and the values of the fitted parameters at different energies are tabulated in Table 1. From this table we can see that the values of $\sigma_H$ and $\sigma_L$ are decreasing with decrease in incident energies of the projectiles. It reflects that as the kinetic energy of projectile will be less and less the hot and cold regions must be smaller and smaller i.e. the thermal equilibrium can be achieved very quickly at low kinetic energy of the projectile.

Table 1: Parameters of the Rayleigh scattering fitting function. The contributions of the Hot and Cold sources are taken to be 0.5 and 0.5, respectively.

| Projectile | Energy (A GeV) | $\sigma_H$ | $\sigma_L$ |
|---|---|---|---|
| $^{84}$Kr | 1.7 | 175 | 96 |
| $^{84}$Kr | 1.5 | 172 | 95 |
| $^{84}$Kr | 0.95-0.80 | 170 | 93 |
| $^{84}$Kr | 0.80-0.50 | 169 | 92 |

Table 2: Derived temperature values for Hot and Cold regions using Maxwell's ideal gas assumptions.

| Projectile | Energy (A GeV) | $T_H$ (MeV) | $T_L$ (MeV) | Ref. |
|---|---|---|---|---|
| $^{8}$B + Em | 1.2 | 10.7 | 1.0 | [11] |
| $^{16}$O + AgBr | 200.0 | 17.7 | 4.2 | [12] |
| $^{24}$Mg + AgBr | 3.7 | 34.7 | 4.8 | [13] |
| $^{84}$Kr + Em | 1.7 | 32.64 | 9.82 | [10] |
| $^{84}$Kr + Em | 1.5 | 31.53 | 9.62 | [4] |
| $^{84}$Kr + Em | 0.95-0.80 | 30.80 | 9.22 | [4] |
| $^{84}$Kr + Em | 0.80-0.50 | 30.44 | 9.02 | [4] |
| $^{197}$Au + Em | 10.6 | 50.00 | 20.00 | [14] |
| $^{208}$Pb + Em | 160.0 | 120.00 | 40.00 | [15] |

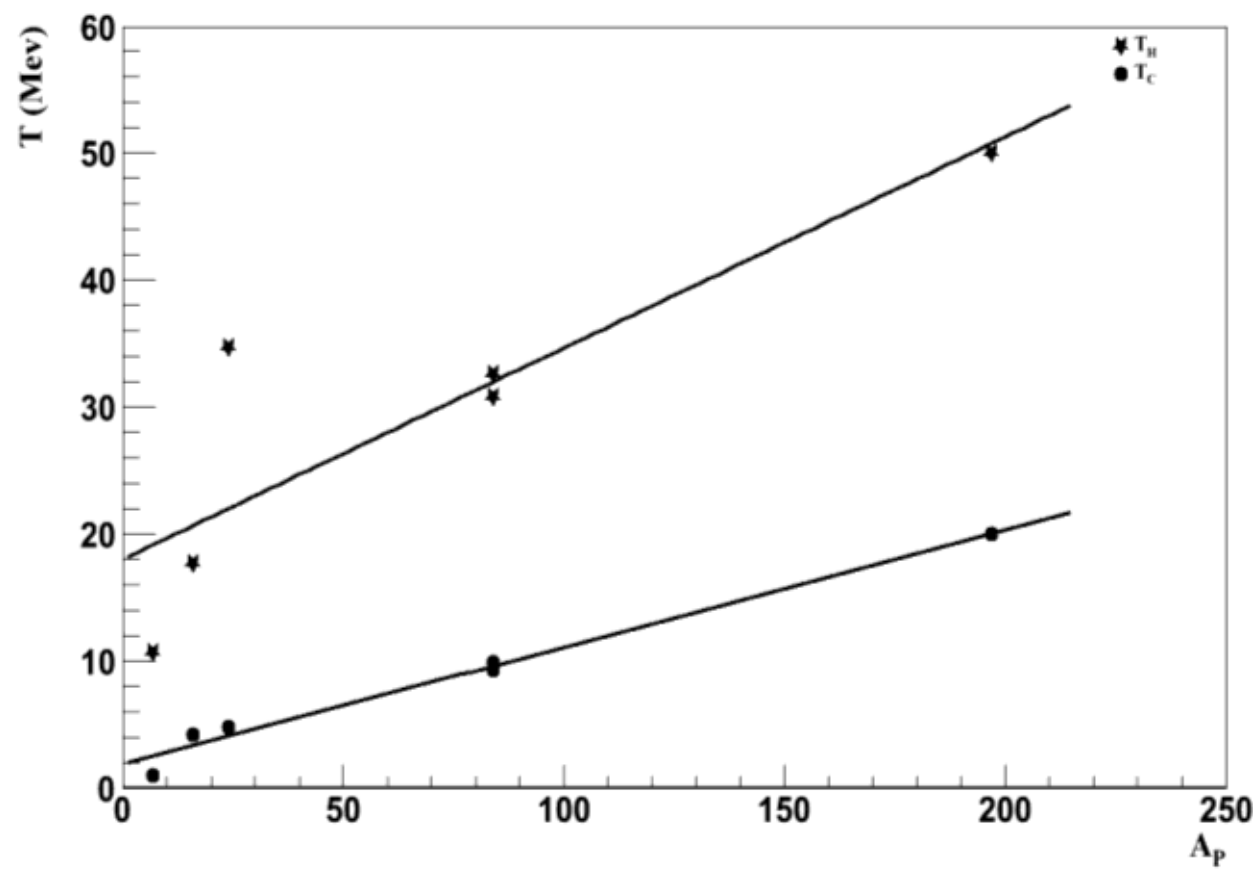

Figure 3: Derived temperature of hot and cold regions is fitted with a best fit straight line function excluding $^{208}$Pb data.

In the frame of the Maxwell's ideal gas model, the corresponding temperatures are also obtained by $\sigma^2/m_p$ for both the sources and tabulated in Table 2 where $m_p$ is the mass of proton. These temperature values are plotted in Figure 2 with respect to the projectile mass number independent of incident energy. The best fit parameter values are 0.092±0.006 (0.166±0.049) and 1.82±0.57 (17.99±4.63) for slope and constant, respectively for cold (hot) region. After excluding the $^{208}$Pb data point the slope and constant values are 0.196±0.021 and 13.16±2.16, respectively with far better $\chi^2$ value. It can be concluded from figure 2 that the hot and cold region temperature has dependence with projectile mass number i.e. size of the colliding system. Therefore it is clear that relativistic fragments produced in high energy nucleus-nucleus collisions can be regarded as the result of a two-source emission. We can conclude that two source model gives a reasonable description in the case of projectile fragments alpha emitted in $^{84}$Kr nuclei interactions with target nuclei at different kinetic energies below 2 GeV per nucleon.

## Acknowledgments

The author (VS) wishes to thank JACoW for their guidance in preparing this template. Authors are also thankful to the accelerator division staff at GSI/SIS and JINR (Dubna) for exposing nuclear emulsion detector with $^{84}$Kr beam and all other assistance. This work is supported by DAE, Mumbai, India.